\documentclass[11pt]{article}
\usepackage{latexsym}  
\usepackage{amssymb}
\usepackage{graphicx}
\usepackage{amsmath}

\newcommand{\be}{\begin{equation}}
\newcommand{\ee}{\end{equation}}
\newcommand{\bea}{\begin{eqnarray}}
\newcommand{\eea}{\end{eqnarray}}
\newcommand{\bref}[1]{(\ref{#1})}
\begin{document}
\begin{titlepage}
\begin{flushright}
\today
\end{flushright}
\vspace{4\baselineskip}
\begin{center}
{\Large\bf  Twenty years after the discovery of $\mu-\tau$ symmetry.}
\end{center}
\vspace{1cm}
\begin{center}
{\large Takeshi Fukuyama
\footnote{E-mail:fukuyama@se.ritsumei.ac.jp}}
\end{center}
\vspace{0.2cm}
\begin{center}

{\small \it Research Center for Nuclear Physics (RCNP), \\ Osaka University, Ibaraki, Osaka, 567-0047, Japan} 
\medskip
\vskip 10mm
\end{center}
\vskip 10mm
\begin{abstract}
It has passed 20 years after we proposed $\mu-\tau$ symmetry in light neutrino mass matrix. This model is simple but reproduced the characterestic properties of lepton sector. After that, during the experimental developments, there have appeared so many extensions but most of those phenomenological models are lacking systematic outlooks towards more fundamental theories. In this paper, we try to consider rather systematic model extensions and application to GUT model.

\end{abstract}
\end{titlepage}
\section{Introduction}
Twenty years ago we proposed first in the world $\mu-\tau$ symmetry in the neutrino mass matrix model \cite{F-N}, 
\begin{equation}
M_\nu=
\left(
\begin{array}{ccc}
0&A&\pm A\\
A&B&C\\
\pm A&C&B
\end{array}
\right).
\label{form}
\end{equation}
in the charged lepton diagonal base. Here $A,B,C$ are real and its components are invariant under $\mu-\tau$ exchange. (1,2) and (1,3) components are equal up to phase convention. This matrix, therefore, has been called $\mu-\tau$ symmetric mass matrix. This leads immediately to $\theta_{23}=\mp\pi/4$ and $\theta_{13}=0$ (double sign in the same order as \bref{form}). This matrix represents the characterestic pattern of the mixing angles which is quite different from that of quark sector. 
The vanishing (1,1) component leads to the small mixing angle (SMA) solution on $\theta_{12}$ (See Eq.\bref{small}), which had survived with large mixing angle (LMA) solution at that time.
However, KamLAND \cite{KamLAND} selected the larger part of solar neutrino angles, and we may set nonzero parameter $D$ in place of the vanishing (1,1) component without breaking $\mu-\tau$ symmetry. In 2013, Daya-Bay\cite{Daya-Bay} made surprise the unexpectedly large $\theta_{13}$. It is impressing that our minimal SO(10) model \cite{FukuSO(10)} discussed in section 4 has suffered from large $\theta_{13}$ before Daya-Bay.

The observed data of leptonic mixing matrix nowaday are summarized as \cite{PDG2016}
\bea
&&\sin^2\theta_{12} = 0.304 \pm 0.014\nonumber\\
&&\Delta m^2_{21} = (7.53 \pm 0.18) \times 10^{-5}~\mbox{eV}^2\nonumber\\
&&\sin^2\theta_{23} = 0.51 \pm 0.05 ~~\mbox{(normal mass hierarchy)}\\
&&\sin^2\theta_{23} = 0.50 \pm 0.05 ~~\mbox{(inverted mass hierarchy)}\nonumber\\
&&\Delta m^2_{32} = (2.44 \pm 0.06) \times 10^{-3}~\mbox{eV}^2~~\mbox{(normal mass hierarchy)}\nonumber\\
&&\Delta m^2_{32} = (2.51 \pm 0.06) \times 10^{-3} \mbox{eV}^2~~\mbox{(inverted mass hierarchy)}\nonumber\\
&&\sin^2\theta_{13} = (2.19 \pm 0.12) \times 10^{-2}\nonumber
\eea
Even in these refined data, our $\mu-\tau$ symmetric model does not lose its significance since such simple and real symmetric model is basically an idealized model and remains valid as the zero'th order approximation of more sophiscated models. 

Indeed there have appeared a vast variety of papers during new experimental developments. Unfortunately, most of those phenomenological models are lacking systematic analyses valid for theoretical developments from phenomenological model to more fundamental one.

In this paper we reconsider $\mu-\tau$ symmetry in these experimental backgrounds and try to fill this deficit.

This paper is organized as follows. In section 2 we review the original $\mu-\tau$ symmetric model.  This model is extended in section 3. In section 4 we argue the correlaton with GUTs. Section 5 is devoted to discussions.

\section{$\mu-\tau$ symmetric model}

Since neutrino oscillation experiments are wholly insensitive to the Majora phases, the Pontecorvo-Maki-Nakagawa-Sakata (PMNS) mixing matrix is in general written in the form
\begin{equation}
U=
\left(
\begin{array}{ccc}
c_{13}c_{12},&c_{13}s_{12},& s_{13}e^{-i\delta}\\
-c_{23}s_{12}-s_{23}c_{12}s_{13}e^{i\delta},&c_{23}c_{12}-s_{23}s_{12}s_{13}e^{i\delta},&s_{23}c_{13}\\
s_{23}s_{12}-c_{23}c_{12}s_{13}e^{i\delta},&-s_{23}c_{12}-c_{23}s_{12}s_{13}e^{i\delta},&c_{23}c_{13}
\end{array}
\right)
\label{mixing}
\equiv
\left(
\begin{array}{ccc}
c_{13}c_{12},&c_{13}s_{12},& s_{13}e^{-i\delta}\\
U_{\mu 1},&U_{\mu 2},&s_{23}c_{13}\\
U_{\tau 1},&U_{\tau 2},&c_{23}c_{13}
\end{array}
\right)
\end{equation}
Here $c_{ij}=\mbox{cos}\theta_{ij},~s_{ij}=\mbox{sin}\theta_{ij}$ as usual. Neutrino mass matrix is written as
\be
M_\nu=U
\left(
\begin{array}{ccc}
-m_1,&0,& 0\\
0,&m_2,&\\
0,&0,&m_3
\end{array}
\right)U^T
\end{equation}
in the charged lepton diagonal base, where we can set $-1$ for $m_1$ using the rephasing. Its explicit components are
\be
\left(M_\nu\right)_{11}=-m_1c_{12}^2c_{13}^2+m_2s_{12}^2c_{13}^2+m_3s_{13}^2e^{-2i\delta},
\ee
\bea
\left(M_\nu\right)_{12}&=&-m_1c_{12}c_{13}U_{\mu 1}+m_2s_{12}c_{13}U_{\mu 2}+m_3s_{13}c_{13}s_{23}e^{-i\delta},\\
\left(M_\nu\right)_{13}&=&-m_1c_{12}c_{13}U_{\tau 1}+m_2s_{12}c_{13}U_{\tau 2}+m_3s_{13}c_{13}c_{23}e^{-i\delta},
\eea
\bea
\left(M_\nu\right)_{22}&=&-m_1U_{\mu 1}^2+m_2U_{\mu 2}^2+m_3c_{13}^2s_{23}^2,\\
\left(M_\nu\right)_{33}&=&-m_1U_{\tau 1}^2+m_2U_{\tau 2}^2+m_3c_{13}^2c_{23}^2.
\eea
As is easily checked, $\theta_{23}=\pm \pi/4$ and $\theta_{13}=0$ is the unique solutions for real $\left(M_\nu\right)_{12}=\mp \left(M_\nu\right)_{13}$, $(M_\nu)_{22}=(M_\nu)_{33}$ relations (double sign corresponds). 

If we adopted $\theta_{23}=\pi/4$ and $\theta_{13}=0$, then neutrino mass matrix becomes
\begin{eqnarray}
M_\nu&=&U
\left(
\begin{array}{ccc}
-m_1&0&0\\
0&m_2&0\\
0&0&m_3
\end{array}
\right)
U^T\nonumber\\
&=&\left(
\begin{array}{ccc}
-c_1^2m_1+s_1^2m_2&{1\over\sqrt{2}}c_1s_1(m_1+m_2)&-{1\over\sqrt{2}}c_1s_1(m_1+m_2)\\
{1\over\sqrt{2}}c_1s_1(m_1+m_2)&{1\over 2}(-s_1^2m_1+c_1^2m_2+m_3)&{1\over 
2}(s_1^2m_1-c_1^2m_2+m_3)\\
-{1\over\sqrt{2}}c_1s_1(m_1+m_2)&{1\over 2}(s_1^2m_1-c_1^2m_2+m_3)&{1\over 
2}(-s_1^2m_1+c_1^2m_2+m_3)
\end{array}
\right),\nonumber\\
\label{mixing2}
\end{eqnarray}
where $c_1 (s_1) \equiv \cos\theta_{12}(\sin\theta_{12})$ for brevity.
In \bref{mixing2} we assumed further
\be
-c_{12}^2m_1+s_{12}^2m_2=0,
\label{small}
\ee
and we obtain the mass matrix of \bref{form}. Neutrino masses are expressed in terms of $A,B,C$,
\bea
\label{massform1}
-m_1&=&\frac{1}{2}\left[B\pm C-\sqrt{8A^2+(B\pm C)^2}\right],\nonumber\\
m_2&=&\frac{1}{2}\left[B\pm C+\sqrt{8A^2+(B\pm C)^2}\right],\\
m_3&=&B\mp C.\nonumber
\eea
The double sign corresponds to \bref{form}. 
Eq.\bref{small} indicates that neutrinoless double beta decay does not happen in this limit, since
\be
\langle m\rangle_{ee} \equiv \sum_i U_{ei}m_iU_{ei} =0.
\ee
\bref{small} favored the small mixing angle solution for $\theta_{12}$ which still had survived at that time.
Also the vanishing (1,1) component is interested in connection with seesaw invariant mass matrix \cite{F-N}.
\be
(a)=\left(
\begin{array}{ccc}
0&A&A\\
A&B&C\\
A&C&C
\end{array}
\right)~~(b)=\left(
\begin{array}{ccc}
0&A&A\\
A&B&B\\
A&B&C
\end{array}
\right)~~(c)=\left(
\begin{array}{ccc}
0&A&A\\
A&B&C\\
A&C&B
\end{array}
\right)~~(d)=\left(
\begin{array}{ccc}
0&A&-A\\
A&B&C\\
-A&C&B
\end{array}
\right)
\ee
(c) is transformed to (d) by the interchange of $C$ to $-C$ and these are physically equivalent as follows. If we leave $\theta_{23}$ as a free parameter and keep the assumtion \bref{small}, then $M_\nu$ is reduced to
\be
\left(
\begin{array}{ccc}
0&c_2\sqrt{m_1m_2}&-s_2\sqrt{m_1m_2}\\
c_2\sqrt{m_1m_2}&(-m_1+m_2)c_2^2+m_3s_2^2&(m_1-m_2+m_3)c_2s_2\\
-s_2\sqrt{m_1m_2}&(m_1-m_2+m_3)c_2s_2&(-m_1+m_2)s_2^2+m_3c_2^2
\end{array}
\right),
\ee
where $c_2 (s_2) \equiv \cos \theta_{23} (\sin \theta_{23})$ for brevity.
Therefore, (c) and (d) are corresponding to $s_{23}=-\pi/4$ and $\pi/4$, repecetively. $\theta_{23}$ has been determined from the mixing factor $\sin^22\theta_{23}$
and they are equivalent.
(a) and (b) are also substantially same and from Eq.\bref{mixing2} they are enforced to $m_3\approx 0$. This is the case of inverted hierarchy. 
This symmetric and seesaw invariant concepts is extended to two-zero texture \cite{Zhou}.

 Eq.\bref{small} was imposed as it enables us to fix all three masses by the same three parameters as \bref{massform1}. You can easily generalize this simple model as
\begin{equation}
M_{\nu}'=
\left(
\begin{array}{ccc}
D&A&\pm A\\
A&B&C\\
\pm A&C&B
\end{array}
\right).
\end{equation}

Eqs.\bref{small} and \bref{massform1} in this case are generalized to
\be
\tan 2\theta_{12}=\frac{2\sqrt{2}A}{B\pm C-D},
\label{tan}
\ee
\bea
-m_1&=&\frac{1}{2}\left[B\pm C+D-\sqrt{8A^2+(B\pm C-D)^2}\right],\nonumber\\
m_2&=&\frac{1}{2}\left[B\pm C+D+\sqrt{8A^2+(B\pm C-D)^2}\right],\\
m_3&=&B\mp C. \nonumber
\label{massform2}
\eea
If $B\pm C-D=A, ~U$ goes to tri-bi-maximal case \cite{TBM},
\be
U_{TB}=
\left(
\begin{array}{ccc}
\frac{2}{\sqrt{6}}&\frac{1}{\sqrt{3}}&0\\
-\frac{1}{\sqrt{6}}&\frac{1}{\sqrt{3}}&-\frac{1}{\sqrt{2}}\\
-\frac{2}{\sqrt{6}}&\frac{1}{\sqrt{3}}&\frac{1}{\sqrt{2}}
\end{array}
\right).
\label{tri}
\ee

\section{Extension of $\mu-\tau$ symmetric model}
The original $\mu-\tau$ symmetric model was real and can not involve CP phases.
So a naive extension is to extend it to a complex and symmetric mass matrix retaining $\mu-\tau$ symmetry. The reasoning why we adhere to a symmetric matrix will be explained in section 4.
Eq.\bref{form} is a real symmetric matrix. Its naive extension is 
\be
M_\nu=
\left(
\begin{array}{ccc}
D&A&\pm A^*\\
A&B&C\\
\pm A^*&C&B^*
\end{array}
\right)~~(\mbox{double sign corresponds}),
\label{cform}
\ee
where $D,C\in \mathcal{R},~A,B\in \mathcal{C}$. This matrix is diagonalized by the unitary matrix \cite{Harrison},
\be
U=
\left(
\begin{array}{ccc}
u_1&u_2&u_3\\
v_1&v_2&v_3\\
\pm v_1^*&\pm v_2^*&\pm v_3^*
\end{array}
\right) ~~(\mbox{double sign correspond to that of \bref{cform}})
\label{HPunitary}
\ee
as
\be
M_\nu M_\nu^\dagger = UD_{\nu}^2U^\dagger.
\ee

Then
\be
M_\nu M_\nu^\dagger=UD_\nu^2 U^\dagger=
\left(
\begin{array}{ccc}
z&w&\pm w^*\\
w^*&x&y\\
\pm w&y^*&x
\end{array}
\right).
\label{HPmass}
\ee 
In Eqs. \bref{cform}, \bref{HPunitary}, and \bref{HPmass}, double sign corresponds. Here $x,z\in\mathcal{R},~y,w\in\mathcal{C}$.
In this complex form we have, in addition to $\theta_{23}=\frac{\pi}{4},~\theta_{13}=0$, another solution, 
\bea
&&\theta_{23}=\frac{\pi}{4}, ~\delta=\pm \frac{\pi}{2}~~\mbox{for minus signature}\\
&&\theta_{23}=-\frac{\pi}{4},~ \delta=\pm \frac{\pi}{2}~~\mbox{for plus signature}
\eea
$\delta=-\frac{\pi}{2}$ is interesting since it is the global minimum (though $1~\sigma$) \cite{Capozzi:2013csa}.

One strategy for extending $\mu-\tau$ symmetry is the following: The extensions  not only explain the leptonic CP phase but also must include quark sector. This is because we are considering GUT as its more fundamental final correspondent. One of such examples preserves $\mu-\tau$ symmetry up to phase but breaks the symmetric property of mass matrices \cite{Fuku1} like,
\be
M_{f}=P_f^\dagger\hat{M_f}P_f
\equiv
P_f^\dagger
\left(
\begin{array}{ccc}
D_f&A_f&A_f\\
A'_f&B_f&C_f\\
A'_f&C_f&B_f
\end{array}
\right)P_f
\label{massform3}
\ee
where
\be
P=\mbox{diag}(e^{i\alpha_{f1}}, e^{i\beta_{f2}},e^{i\gamma_{f3}})
\ee
and $f$ includes up-type and down-type quark.
First we diagonalize $\hat{M_f}$ by two orthogonal matrices $O_{f1}$ and $O_{f2}$ as
\be
O_{f1}^T\hat{M}_fO_{f2}=\mbox{diag}(m_{f1}, m_{f2}, m_{f3})
\ee
\be
O_{fi}=U_{TB}
\left(
\begin{array}{ccc}
\cos \varphi_{fi}&-\sin \varphi_{fi}&0\\
\sin \varphi_{fi}&\cos \varphi_{fi}&0\\
0&0&1
\end{array}
\right)~~(i=1,2)
\label{TBMM}
\ee
where $U_{TB}$ is a tri-bi-maximal mixing matrix \bref{tri}.
We found that these matrices are consistent with the experimental data of CKM
mixing matrix. This is the extension to quark sector but is left on the same phenomenological level as the original work of lepton sector. Hereafter we restrict ourselves in symmetric mass matrices again. If we involve quark sector, it must reveal some higher symmetric (more fundamental) new character. In this sence, though there are a vast variety of these extensions, most of these phenomenological extensions have no systematic idea leading to more fundamental theoretical models. 
For the route from (low energy) phenomenological model to (high energy) more fundamental one, some symmetry must play an important role. 

Let us explain it in the well known example:
QED lagrangian has $U(1)$ and Lorentz invariances,
\be
L^{QED}=-\frac{1}{4}F_{\mu\nu}F^{\mu\nu}+\frac{\theta}{16\pi^2}\epsilon^{\mu\nu\rho\sigma}F_{\mu\nu}F_{\rho\sigma}+L_{matter}.
\label{QED}
\ee
Under T transformation, ${\bf E}\to-{\bf E}$ and ${\bf B}\to{\bf B}$, and therefore $\theta\to -\theta$, if we preserve T invariance. In another word, T invariance requires $\theta=0$.
Lorentz invariance breaks to spatial rotation invariance for dielctics,
\be
L^{dielectric}=\frac{\epsilon}{2}{\bf E}^2-\frac{1}{2\mu}{\bf B}^2+\frac{\theta}{4\pi^2}{\bf E}\cdot{\bf B}.
\label{dielectric}
\ee
Thus, permeability ($\mu$) and permittivity ($\epsilon$) characterize this symmetry breaking. Eq.\bref{dielectric} (\bref{QED}) may be considered as a phenomenological (a fundamental) model. According to this general idea, how the phenomenological $\mu-\tau$ symmetry is incorporated to higher symmetry group or more comprehensive model ?  It is natural to incorporate quark sector in this higher symmetric world. We consider here $A_4$ group as a candidate for it \cite{Fuku4}.
$A_4$ is the four degreed symmetry group with even permutation whose elements we denote
as $(a_1, a_2, a_3, a_4) \cite{Altarelli}.~ A_4$ is generated by the $S$ and $T$ and their products, which satisfy
\be
S^2 = T^3 = (ST)^3 = 1. 
\ee
The three-dimensional unitary representation, in a basis where the element $S$ is diagonal,
is built up from:
\be
S =
\left(
\begin{array}{ccc}
1& 0& 0\\
0& -1& 0\\
0& 0& -1
\end{array}
\right),~~
T =
\left(
\begin{array}{ccc}
0& 1& 0\\
0& 0 &1\\
1& 0& 0
\end{array}
\right).
\ee
Let us practice the transformation, for instance, $ST$ to $V~\equiv~(a_1, a_2, a_3)^T$
\be
TV =
\left(
\begin{array}{c}
a_2\\
a_3\\
a_1
\end{array}
\right),~~
STV =
\left(
\begin{array}{c}
a_2\\
-a_3\\
-a_1
\end{array}
\right).
\ee
The rule of the game for reading the permutation group of four degree from three dimensional
vector is to make plus element change to $a_4$ and do minus signs interchange, and
$ST$ corresponds to $(a_4, a_1, a_3, a_2)$. Thus $S$ means the $2 \leftrightarrow 3$ symmetry and $T$ does cyclic permutation or equivalently $Z_3$.
Mathematically this is the elementary example of Sylow's theorem \cite{Sylow}. The order of
$A_4$ is $12 = 2^2\times 3$, and it is the product of normal subgroup $V_4$, composed of $(1, 2)(3, 4),~ (1, 3)(2, 4),~ (1, 4)(2, 3), ~1$ and $Z_3$. Thus $A_4 \supset (2 \leftrightarrow 3) ~\mbox{symmetry}~\times ~Z_3$. Namely, $\mu-\tau$ symmetry may be considered the residual symmetry broken from $A_4$. This fact is very important for the model buiding of more fundamental theories. 
So far we have not considered $Z_3$, and let us consider how $Z_3$ symmetry appears. Corresponding to this extension, we generalize from the lepton sector to the quark-lepton sector, denoting their fields as $\psi_i$ ($i$: generation), and call 2-3 symmetry instead of $\mu-\tau$ summetry in that case. 

We assign $Z_3$ charge of each generation of fermions so as to be compatible with 2-3 symmetry \cite{Koide},
\be
\psi_{1L}~\to~\psi_{1L},~\psi_{2L}~\to~\omega\psi_{2L},~\psi_{3L}~\to~\omega\psi_{3L},
\ee
where $\omega^3=+1$. Then, the bilinear terms $\overline{q}_{Li}u_{Rj},~\overline{q}_{Li}d_{Rj},~\overline{l}_{Li}\nu_{Rj},~\overline{l}_{Li}e_{Rj}$ and $\overline{\nu_{Ri^c}}\nu_{Rj}$ are transformed as follows:
\be
\left(
\begin{array}{ccc}
1&\omega^2&\omega^2\\
\omega^2&\omega&\omega\\
\omega^2&\omega&\omega
\end{array}
\right),
\ee
where
\be
q_L=\left(
\begin{array}{c}
u_L\\
d_L
\end{array}
\right).~~l_L=\left(
\begin{array}{c}
\nu_L\\
e_L
\end{array}
\right).
\ee
Therefore, if we assume two SU(2) doublet Higgs scalars $H_1$ and $H_2$, which are transformed as
\be
H_1 \to \omega H_1,~~ H_2 \to \omega^2H_2,
\ee
the Yukawa interactions are given as follows
\bea
H_{Yukawa}&=&\sum_{A=1,2}\left(Y_{(A)ij}^u\overline{q}_{Li}\tilde{H}_Au_{Rj}+Y_{(A)ij}^d\overline{q}_{Li}H_Ad_{Rj}\right)\nonumber\\
&+&\sum_{A=1,2}\left(Y_{(A)ij}^\nu\overline{l}_{Li}\tilde{H}_A\nu_{Rj}+Y_{(A)ij}^e\overline{l}_{Li}H_Ae_{Rj}\right)\\
&+&\left(Y_{(1)ij}^R\overline{\nu_{Rj}^c}\tilde{\Phi^0}\nu_{Rj}+Y_{2ij}^e\overline{\nu_{Rj}^c}\Phi^0\nu_{Rj}\right)+h.c.,\nonumber
\eea
where
\be
H_A=\left(
\begin{array}{c}
H_A^+\\
H_A^0
\end{array}
\right), ~~\tilde{H}_A=\left(
\begin{array}{c}
\overline{H}_A^+\\
-H_A^-
\end{array}
\right).
\ee
Therefore,
\be
Y_{(2)}^u,~Y_{(1)}^d,~Y_{(2)}^\nu,~Y_{(1)}^e,~Y_{(1)}^R=\left(
\begin{array}{ccc}
0&0&0\\
0&*&*\\
0&*&*
\end{array}
\right),~~
Y_{(1)}^u,~Y_{(2)}^d,~Y_{(1)}^\nu,~Y_{(2)}^e,~Y_{(2)}^R=\left(
\begin{array}{ccc}
0&*&*\\
*&0&0\\
*&0&0
\end{array}
\right)
\label{texture0}
\ee
In \bref{texture0}, the symbol $*$ denotes non-zero quantities. Here, in order to give heavy Majorana masses of the right-handed neutrinos $\nu_R$, we have assumed an SU(2) singlet Higgs scalar $\Phi^0$, which is transformed as $H_1$. Mass matrices are sums of $Y_{(1)}$ and $Y_{(2)}$ and their (1,1)
element must be vanished:
\be
\left(
\begin{array}{ccc}
0&*&*\\
*&*&*\\
*&*&*
\end{array}
\right).
\ee
Thus zero-texture model becomes another useful character as well as symmetric property. Then how far can we go along this line of thought ? In this case, neutrino mass matrix may have the special property of seesaw mechanism \cite{Yanagida}, and the concept of the seesaw invariance plays an important role \cite{F-N, Zhou}. Two-zero texture is interesting from the parameter counting. 
Mass matrix $M_\nu$ is determined by 9 out-put parameters, 3 masses, 3 angles and 3 CP phases (one Dirac $\delta$ and 2 Majorana phases) in the charged lepton flavour diagonal base. Two-zero texture gives four constraints and 9-4=5 in-put free parameters \cite{Fritzsch, Ramond, Xing}. 
Among others, the following textures are very important,
\begin{equation}
M_{\nu}^{(1)}=
\left(
\begin{array}{ccc}
0&*&0\\
*&*&*\\
0&*&*
\end{array}
\right)~~
\mbox{and}~~M_{\nu}^{(2)}=\left(
\begin{array}{ccc}
0&0&*\\
0&*&*\\
*&*&*
\end{array}
\right).
\label{texture}
\end{equation}
They are related by $\mu-\tau$ excahange, $M_{\nu}^{(2)}=P_{23}M_{\nu}^{(1)}P_{23}$. Here 
\begin{equation}
P_{23}=
\left(
\begin{array}{ccc}
1&0&0\\
0&0&1\\
0&1&0
\end{array}
\right).
\end{equation}
As mentioned above, five parameters remain free in two-zero texture model. Therefore, if $\theta_{ij},~\Delta m_{sol}^2,~\Delta m_{atm}^2$ are determined, $\delta,~\rho,~\sigma$ are predicted.
\bea
&&m_1\approx s_{13}t_{23}t_{12}m_3e^{i\delta} \nonumber\\
&&m_2\approx s_{12}t_{23}/t_{12}m_3e^{i\delta}\\
&&\Delta m^2_{solar}/\Delta m^2_{atm}\approx s_{13}^2t_{23}^2|t_{12}^2-1/t_{12}^2|
\eea
for $M_\nu^{(2)}$ case \cite{Xing}, where $t_{ij}\equiv \tan\theta_{ij}$.
For $M_\nu^{(1)}$ case is obtained by replacing $t_{23}$ by $-1/t_{23}$.
Thus $M_{\nu}^{(1)}$ and $M_{\nu}^{(2)}$ give similar results.
This fact is used in GUT formulation as as will be discussed in the next section.

\section{Mass matrix model and GUT}
GUT models basically search the vertical structure of quark-lepton of one generation. Inter-family (horizontal) relations like Yukawa structure are not predicted. $\mu-\tau$ symmetry may be  clue to this extension.

In the previous section, we considered that $2\leftrightarrow 3$ symmetry and $Z_3$ suggest zero texture solution. So we consider GUT implimented witbh texture. For that purpose we set two requirements

\begin{itemize}
\item
GUT model itself must have few ambiguities and be predictive enough.
\item
The reliable mass texture model should be adopted.
\end{itemize}

The $SO(10)$ grand unified theory (GUT) can provide the most promising framework to unify quarks and leptons, because the entire SM matter contents of each generation (including a right-handed neutrino)
can be unified in a single irreducible representation, $\bf 16$.
A particular attention has been paid to
the renormalizable minimal $SO(10)$ model, 
where two Higgs multiplets 
 $\{{\bf 10} \oplus {\bf \overline{126}}\}$ 
 are utilized for the Yukawa couplings with the matter representation \cite{Babu:1992ia}. 
The couplings to the {\bf 10} and $\overline{\bf 126}$ Higgs fields
can reproduce realistic charged fermion and neutrino mass matrices using their phases thoroughly \cite{Matsuda:2000zp, Fukuyama:2002ch}. $\overline{\bf 126}$ Higgs is selected since it includes $({\bf 10,~1,~3})$ and $({\bf \overline{10},~1,~3})$ under the Pati-Salam subgroup which induce type I and type II seesaw mechanism, respectively. Yukawa coupling is given by
\begin{eqnarray}
 W_Y = Y_{10}^{ij} {\bf 16}_i H_{10} {\bf 16}_j 
           +Y_{126}^{ij} {\bf 16}_i H_{126} {\bf 16}_j \; , 
\label{Yukawa1}
\end{eqnarray} 
where ${\bf 16}_i$ is the matter multiplet of the $i$-th generation,  
 $H_{10}$ and $H_{126}$ are the Higgs multiplet 
 of {\bf 10} and $\overline{\bf 126} $ representations. 

 The Yukawa coupling, after SO(10) symmetry is broken down to the standard model, is given as follows:
\begin{eqnarray}
 \label{massmatrix}
  M_u &=& c_{10} M_{10} + c_{126} M_{126}   \nonumber \\
  M_d &=&     M_{10} +     M_{126}   \nonumber \\
  M_D &=& c_{10} M_{10} -3 c_{126} M_{126}    \\
  M_e &=&     M_{10} -3     M_{126}   \nonumber \\
  M_T &=& c_T M_{126} \nonumber \\  
  M_R &=& c_R M_{126}  \nonumber \; . 
\end{eqnarray}
Here $c_{10},~c_{126},~c_T,~c_R$ are comlex constants. It should be remarked that mass matrices are complex and symmetric matrices because of the group property of ${\bf 10, \overline{126}}$ representations.
Here we proceed to incorporate two-zero texture i n this model. We adopted the two-zero texture mass $M_\nu^{(2)}$ of \bref{texture}.

Unfortunately, the data fittings have been performed by inputting quark sector spectrum and outputted the lepton sector. In this approach, neutrino mass texture is contaminated by the special base adopted in quark sector and shows no clear texture in $M_\nu$. The reason why we adopted the quark sector as input data is that the leptoin sector had been more ambiguous than the quark sector. Nowaday, however, the situation changed. The lepton parameters are more accurate rather than the quark ones, and a large threshold correction is expected in the
quark sector in SUSY models. In that sence, it is better to perform the fitting by inputting the parameters in the lepton sector. The formulation is presented in not only such a practical purpose, but also to make clear the property of the solution with $v_R\approx 10^{16}$ GeV. Here $v_R$ is the typical intermediate enrygy scale, and usualy adopted $10^{13}$ GeV spoils the gauge coupling unifications \cite{Malinsky}. Real data fitting revealed that in $v_R\approx 10^{16}$ GeV solution, the down quark mass is smaller than the observation, and (1,1) and (1,2) elements of $(M_\nu)$ are smaller than the other elements in the fit result.
The deficit of down quark mass can be considered as the threshold correction.
Under the assumtion of 
$(M_\nu)_{11}=(M_\nu)_{12}$ lead us to \cite{Fuku3}
\begin{equation}
\cos\delta_{\rm PMNS} = 
\frac{ \frac{\Delta m_{\rm sol}^2}{\Delta m_{\rm atm}^2} \cos2\theta_{13}\sin^22\theta_{12}   - 
4\sin^2\theta_{13} \left(\frac{\Delta m_{\rm sol}^2}{\Delta m_{\rm atm}^2} \cos^4\theta_{12}+\cos2\theta_{12}\right)
\tan^2\theta_{23}}{4\sin^3\theta_{13} \left(1+  \frac{\Delta m_{\rm sol}^2}{\Delta m_{\rm atm}^2} \cos^2\theta_{12}\right)\sin2\theta_{12}\tan\theta_{23}}.
\label{texture2}
\end{equation}
In Fig.1 we plot the relation between $\delta_{PMNS}$ and $\theta_{23}$ in the assumption. Of course those two mass matrix elements are not exactly zero in the fits, and provides a guide to understand the fit results for the prediction of the PMNS phase depending on the mixing angle. In Fig.2, we show the plot of proton decay, $\tau(p\to K\overline{\nu})$, in the $\delta_{PMNS}-\theta_{23}$ plane. As expected , the partial proton lifetime is larger near the curve of zero-texture in Fig.1. Near the curve, the lifetime is about 10 times bigger than the current experimental bound $\tau(p\to K\overline{\nu}) >0.59\times 10^{34}$ years. Please see \cite{Fuku3,Fuku5} for the detail. 

\begin{figure}[h]
\begin{center}
\includegraphics[width=0.5\textwidth]{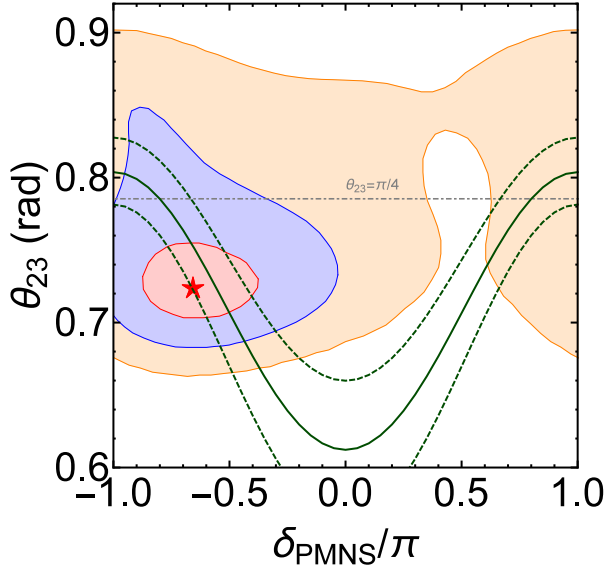}
\caption{\sl \small 
The plot from the relation in Eq.\bref{texture2}. Along the green line, the (1,1) band (1,2) elements can become zero in the neutrino masss matrix in the basis where the charged lepton mass matrix diagonal. The dotted lines are drawn using $3\sigma$ range of the mixing parameters. We also overlap the current $1\sigma$ (red), $2\sigma$ (blue), $3\sigma$ (orange) region of the global analysis in Ref. \cite{Capozzi:2013csa}. The star symbol shows the current best fit of the global analysis. This figure is cited from \cite{Fuku5}. (For interpretation of the references to color in this figure legend, the reader is referred to the web version of this article.)
}
\end{center}
\label{fig:PMNS}
\end{figure}

\begin{figure}[h]
\begin{center}
\includegraphics[width=0.5\textwidth]{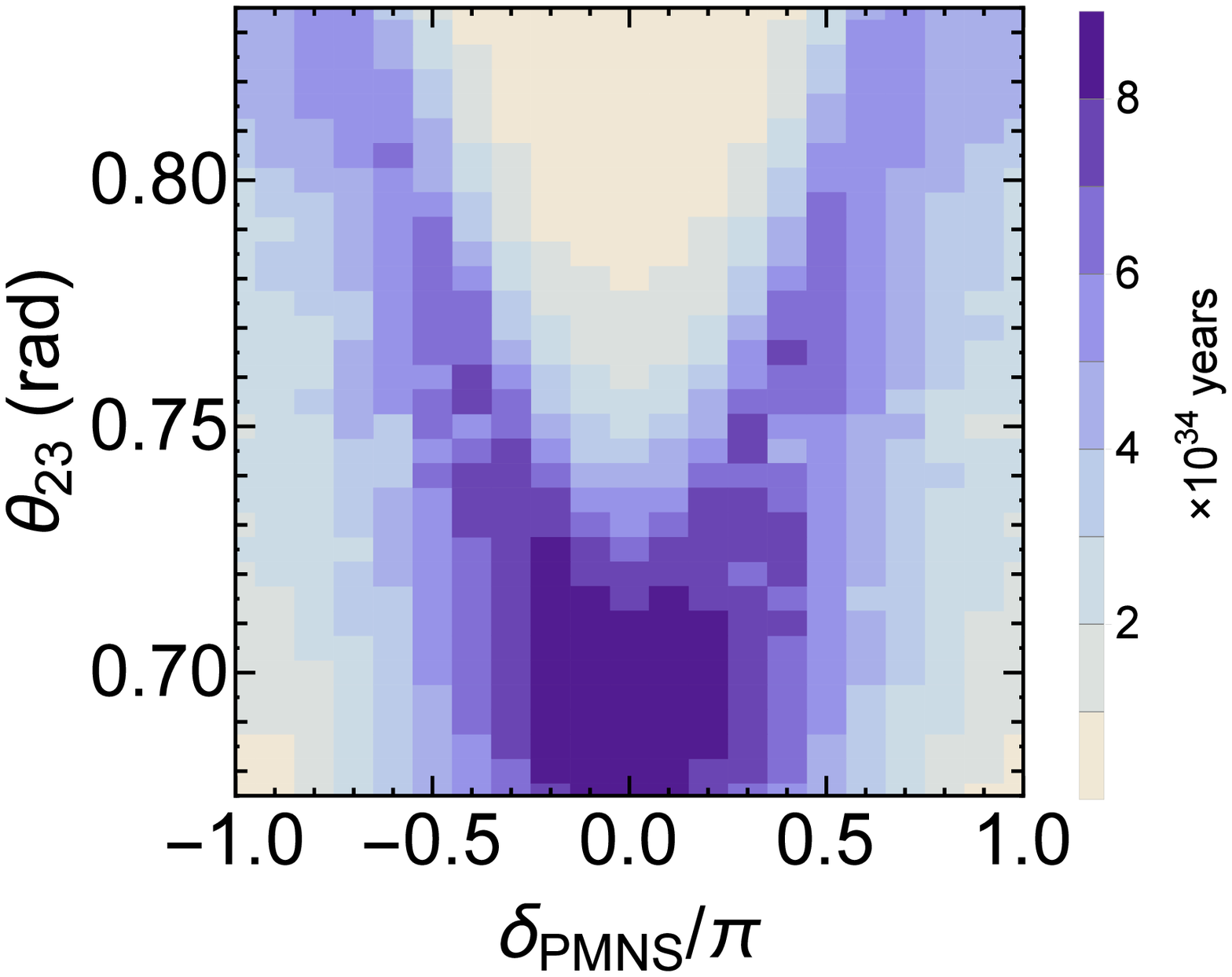}
\caption{Numerical result of the partial lifetime of the $p\to K\overline{\nu}$ decay. The lifetime is larger near the curve given by $M_\nu^{(2)}$ of \bref{texture}.
$\tau(p\to K\bar\nu)^{\rm EXP} > 0.59 \times 10^{34}$ years \cite{Abe}. This figure is cited from \cite{Fuku5}.
}
\end{center}
\label{run2}
\end{figure}

\section{Discussion}
We have not connected $2-3$ symmetry or $A_4$ symmetry directly with GUT symmetry SO(10). Naively it seems to be natural to consider SO(10)$\times A_4$ symmetry.
However, the merit of renormalizable minimum SO(10) GUT is that all mass matrices, as you see in \bref{massmatrix}, are represented by only two mass matrices $M_{10}$ and $M_{126}$. So it makes this model very predictive and few room to add any assumption. If we incorporate $A_4$ into SO(10) GUT naively, it brings about many ambiguities on how to specify $A_4$ to Higgs and inflavons et.al. \cite{King}. This spoils the high predictivity of the minimal SO(10) GUT.
On the other hand, two-zero texture is very useful for full data fitting scan because we know $\left(M_\nu\right)_{11}\approx\left(M_\nu\right)_{12}\approx 0$ phenomenologically. So we can scan around this neighbourhood. And indeed we have found very good fitting around this solution. It is very interesting that long proton decay is obtained along this solution like Fig.2. Such collaboration of GUT (fundamental theory) with phenomenological model is unprecedented. $2-3$ symmetry and two-zero texture make clear the GUT solution as well as the practical usefulness of comprehensive data fittings. Thus we can not only fix all mass matrices but also predict many unobserved parameters, like proton decay and lepton flavour violation. Moreover, using the SUSY breaking boundary condition indicateded in \cite{Fuku3}, we can fit the other almost all known bounds like LFV and $\Omega_{DM}$ BR$(B_s^0\to \mu^+\mu^-)$ etc. \cite{FOT}

\section*{Acknowledgments}
This paper is based on a talk given in a mini-workshop on gquarks, leptons and family gauge bosonshDecember 26-27, 2016 Osaka. The author expresses his sincere thanks to the organizer Y.Koide. He is grateful to H.Nishiura for his excellent collaboration on $\mu-\tau$ symmetry. He is also deeply indebted to Y.Mimura and K.Ichikawa in the recent works on the SO(10) GUT. This work is supported in part by Grant-in-Aid for Science Research from Japan Ministry of Education, Science and Culture (No.~26247036).


\newpage
\begin{thebibliography}{99}

\bibitem{F-N}
T.Fukuyama and H.Nishiura, Proceeding of 1997 Shizuoka Workshop on Masses and Mixings of Quarks and Leptons, World Scientific Pub. Comp. (1997),
hep-ph/9702253,
\bibitem{KamLAND}
K.Eguchi et al. [KamLAND Collaboration], Phys.Rev.Lett.{\bf 90}, 021802 (2003).
\bibitem{Daya-Bay}
F.P.An et al. (Daya-Bay Collaboration), Phys.Rev.Lett. {\bf 108}, 171803 (2013). 
\bibitem{FukuSO(10)}
See for review, T.Fukuyama, Int.J.Mod.Phys. A{\bf 28}, 1330008 (2013). 
\bibitem{PDG2016}
C. Patrignani et al. (Particle Data Group), Chin. Phys. C{\bf 40}, 100001 (2016).
\bibitem{Zhou}
Z.z.Xing and S.Zhou, Phys.Lett. B{\bf 606}, 145 (2005).
\bibitem{TBM}
P.F.Harrison, D.H.Perkins, and W.G.Scott, Phys.Rev.Lett. B{\bf 530} 167 (2002).
\bibitem{Harrison}
P.F.Harrison and W.G.Scott, Phys.Lett. B{\bf 547} 219 (2002).
\bibitem{Capozzi:2013csa}
F.Capozzi, G.L.Fogli, E.Lisi, A.Marrone, D.Montanino,A.Palazzo, Phys.Rev. D{\bf 89}, 093018 (2014).
\bibitem{Fuku1}
H.Nishiura, K.Matsuda, and T.Fukuyama, Int.J.Mod.Phys.A{\bf 23}, 4557 (2008).
\bibitem{Fuku4}
T.Fukuyama, arXiv:0804.2107[hep-ph]
\bibitem{Altarelli}
G.Altarelli, Lectures given at the 61'st Scotish Universities Summer School in Physics, St.Andrews, Scotland, 8-23 August 2006 [arXive:hep-ph/0611117];
H. Ishimori et.al., Prog. Theor. Phys. Supplement No.183 (2010).
\bibitem{Sylow}
See, for instance, B.Huppert, Endliche Gruppen I (Springer, 2'nd ed. 1983).
\bibitem{Koide}
Y.Koide, H.Nishiura, K.Matsuda, T.Kikuchi, and T. Fukuyama, Phys.Rev.D{\bf 66}, 093006 (2002).
\bibitem{Yanagida}
P. Minkowski, Phys. Lett. B67 (1977) 421; T. Yanagida in Workshop on Unified Theories, KEK Report 79-18, p. 95, 1979. Conf. Proc. C 7902131, 95 (1979); Prog. Theor. Phys. {\bf 64}, 1103 (1980); M. Gell-Mann, P. Ramond and R. Slansky,
Supergravity, p. 315. Amsterdam: North Holland, 1979; S. L. Glashow, 1979 Cargese 42 Summer Institute on Quarks and Leptons, p. 687. New York: Plenum, 1980; R. N.Mohapatra and G. Senjanovic, Phys. Rev. Lett. {\bf 44}, 912 (1980).
\bibitem{Fritzsch}
H. Fritzsch, Phys.Lett. B{\bf 73}, 317 (1977).
\bibitem{Ramond}
P.Ramond, R.G.Roberts, and G.G.Ross, Nucl.Phys. B{\bf 406}, 9 (1993)
\bibitem{Xing}
P.H.Frampton, S.L.Glashow, D.Marfatia, Phys.Lett. B{\bf 536}, 79 (2002);
H.Fritzsch, Z.Z.Xing, and S.Zhou, JHEP, {\bf 09}, 083 (2011)

\bibitem{Babu:1992ia}
  K.~S.~Babu and R.~N.~Mohapatra,
  Phys. Rev. Lett.  {\bf 70}, 2845 (1993)
  [hep-ph/9209215].
\bibitem{Matsuda:2000zp} 
  K.~Matsuda, Y.~Koide and T.~Fukuyama,
  Phys.\ Rev.\ D {\bf 64}, 053015 (2001)
  [hep-ph/0010026];
%
  K.~Matsuda, Y.~Koide, T.~Fukuyama and H.~Nishiura,
  Phys.\ Rev.\ D {\bf 65}, 033008 (2002)
  [Erratum-ibid.\ D {\bf 65}, 079904 (2002)]
  [hep-ph/0108202].

 \bibitem{Fukuyama:2002ch} 
  T.~Fukuyama and N.~Okada,
  JHEP {\bf 0211}, 011 (2002)
  [hep-ph/0205066].
\bibitem{Malinsky} 
  S.~Bertolini, T.~Schwetz and M.~Malinsky,
  Phys.\ Rev.\ D {\bf 73}, 115012 (2006).
\bibitem{Fuku3}
T.Fukuyama, K.Ichikawa and Y.Mimura, Phys.Rev.D{\bf 94}, 075018 (2016).
\bibitem{Fuku5}
T.Fukuyama, K.Ichikawa and Y.Mimura, Phys.Lett. B{\bf 764}, 114 (2017), arXiv:1609.08640[hep-ph].
\bibitem{Abe}
K.Abe et.al., Super-Kamiokande Collaboiration, Phys.Rev.D{\bf 90}, 072005 (2014).
\bibitem{King}
See, for instanvce,
S. Morisi, M. Picariello, E. Torrente-Lujan, Phys.Rev.D{\bf 75}, 075015 (2007);
A.Anttusch, S.F.King, and M.Spinrath, Phys.Rev.D{\bf 83}, 013005 (2011).
\bibitem{FOT}
T.Fukuyama, N.Okada, and H.M.Tran, Sparticle spectroscopy of the minimal SO(10) model. arXiv:1611.08341[hep-ph].

\end{thebibliography}
\end{document}